# IoT アプリケーションの GPU オフロード時の並列処理部抽出とデータ転送回数低減手法

山登庸次† 　野口博史† 　片岡操† 　磯田卓万† 　出水達也†

† NTT ネットワークサービスシステム研究所，東京都武蔵野市緑町 3-9-11
E-mail: †yamato.yoji@lab.ntt.co.jp

**あらまし**　私達は，オープン IoT に向け，ユーザが必要なデータを持つデバイスを動的に発見し，利用する，Tacit Computing 技術とその要素技術として GPU 自動オフロード技術を提案している．しかし，既存技術は，並列処理部の最適化を行うだけで性能改善できるアプリケーションは限られていた．そこで，本稿では，より多くのアプリケーションを改善するため，CPU と GPU 間のデータ転送回数を低減する手法を提案した．私達は，大規模アプリケーションである Darknet に対して適用し，CPU に比べて 3 倍の高速化を確認した．
**キーワード**　オープン IoT, GPGPU，データ転送最適化，遺伝的アルゴリズム，自動オフロード

## Parallel processing area extraction and data transfer number reduction for automatic GPU offloading of IoT applications

Yoji YAMATO†, Hirofumi NOGUCHI†, Misao KATAOKA†, Takuma ISODA†, and Tatsuya DEMIZU†

† Network Service Systems Laboratories, NTT Corporation, 3-9-11, Midori-cho, Musashino-shi, Tokyo
E-mail: †yamato.yoji@lab.ntt.co.jp

**Abstract**　For Open IoT, we have proposed Tacit Computing technology to discover the devices that have data users need on demand and use them dynamically and an automatic GPU offloading technology as an elementary technology of Tacit Computing. However, it can improve limited applications because it only optimizes parallelizable loop statements extraction. Thus, in this paper, to improve performances of more applications automatically, we propose an improved method with reduction of data transfer between CPU and GPU. We evaluate our proposed offloading method by applying it to Darknet and find that it can process it 3 times as quickly as only using CPU.
**Key words**　Open IoT, GPGPU, Data Transfer Optimization, Genetic Algorithm, Automatic Offloading.

## 1. はじめに

近年，IoT（Internet of Things）技術が進展しており（例えば，Industrie 4.0 等 [1]- [3]），デバイス側で収集したデータを OpenStack [4] 上のクラウド（例えば，[5] [6]）を用いて，Spark [7], Storm [8]，MapReduce [9] 等で分析し可視化するといったアプリケーションが出ている．従来 IoT サービスは，デバイスからアプリケーションまで一体構築されたサイロ型が多かった．しかし，コストを下げ多様なサービスを提供するため，デバイスを複数アプリケーションで共有する Open IoT の概念（例えば，[10]）が注目されている．

Open IoT の例として，街中の複数団体が持つカメラを共有し，OpenCV [11] 等を用いて迷子探索やテロリスト発見等，複数の用途に使うことが期待される．しかし，この例で，画像処理を複数の用途で用いることは，どこで分析するとしても，計算リソースが膨大になる課題がある．

一方，近年，IoT 等多彩な分野に対応するため，CPU 以外の計算リソースを用いることが増えている．例えば，GPU（Graphics Processing Unit）を強化したクラウドサーバ [12] で画像処理を行ったり，FPGA（Field Programmable Gate Array）で処理をアクセラレートすることが増えている [13]．

Open IoT では，サービス連携技術等（例えば，[14]- [17]）を用いて，多彩なアプリケーションの創出が期待されるが，更に進歩したハードウェアを生かすことで，高性能化が期待できる．しかし，そのためには，ハードウェアに合わせたプログラミングや設定が必要で，CUDA（Compute Unified Device Archi-

— 1 —

tecture) [18], OpenCL (Open Computing Language) [19] といった技術知識が求められ, ハードルは高い.

そこで, GPU や FPGA を IoT アプリケーションで容易に利用できる様にするため, 動作させる画像処理等の汎用アプリケーションをデプロイする際に, IoT プラットフォームがアプリケーションロジックを分析し, GPU, FPGA に自動で処理をオフロードする事が望まれる.

私達は, 以前に Open IoT 向けプラットフォームとして, 適切なリソースを動的に発見して利用することで, ユーザにパーソナライズしたサービスを構築する Tacit Computing を提案している. 更に, その要素技術として, 動作させるアプリケーションロジックの GPU への自動オフロードを遺伝的アルゴリズムを用いて行う手法を提案している [20]. しかし, [20] では, アプリケーションの適切な並列処理部の自動抽出を主眼としており, 高速化できるアプリケーションがそれほど多くなかった. そこで, 本稿では, より多くのアプリケーションを, 自動で GPU で高性能化することを狙い, GPU へのデータ転送回数低減を備えた技術提案と評価を行う.

## 2. 既存ヘテロハードウェア処理技術

GPU の計算能力を画像処理以外にも使う GPGPU (General Purpose GPU) のための開発環境 CUDA が発展している. CUDA では, C 言語の拡張によるプログラミングを行うが, GPU 等のデバイスと CPU の間のメモリコピー, 解放等を記述する必要があり, 記述の難度は高い.

簡易に GPGPU を行うため, ディレクティブベースで, 並列処理すべき個所を指定し, ディレクティブに従いコンパイラがデバイス向けコードに変換する技術が有る. 技術仕様として, OpenACC[21] 等, コンパイラとして PGI コンパイラ [22] 等がある. また, Java のラムダ式を解釈し GPU にオフロードする Java 環境もある [23].

このように, CUDA, OpenACC 等の技術により, GPU へのオフロードが可能になっている. しかし, GPU 処理は行えるようになっても, 高速化には課題が多い. マルチコア CPU 向けには, 例えば, Intel コンパイラ [24] 等の自動並列化機能を持つコンパイラがある. 自動並列化時は, プログラム上の for 文等の並列可能部を抽出するが, GPU を用いる場合は, CPU-GPU メモリ間のデータ転送オーバヘッドのため性能が出ないことも多い. GPU を用いて流体計算 [25] を高速化する等の例があるが, スキル者が, CUDA でのチューニングや, PGI コンパイラ等で適切な並列処理部を探索すること ([26][27] 等) が必要になっている. [20] では, 遺伝的アルゴリズムを用いることで, GPU 処理で効果のある並列処理部を自動チューニングしていたが, CPU-GPU メモリ間のデータ転送によっては高性能化できないアプリケーションもあった.

このため, スキルが無いユーザが GPU を使ってアプリケーションを高性能化することは難しいし, 自動並列化技術等を使う場合も並列処理可否の試行錯誤や高速化できない場合があった.

私達は, CPU 向けに IoT で汎用的に利用されるアプリケーション (画像処理, 分析処理等) を, ある程度の時間で適切に GPU にオフロードすることを狙う. これにより, ユーザにとっては, 利用する GPU リソースは増えるが, CPU を動かす仮想マシン等のサーバ数を減らし, 結果的にコストダウンすることを実現する.

## 3. GPU 自動オフロード技術の提案

### 3.1 Tacit Computing の GA 用いた GPU オフロード

[20] で述べた Tacit Computing は, ユーザに適切なデバイスを動的に発見し利用する Open IoT コンセプトを実現している. サービス例として, カメラと画像識別プログラムを用いて, 街中のカメラで画像識別し, 徘徊老人が映っているカメラを見つけ, 家族の方に映像を届けるといった例がある. このように動的にサービスを実現した場合は, 利用されるデバイスはサービス構築のタイミングで決まるため, 処理が効率的でない場合が多い. そこで, Tacit Computing では, 画像分析等の処理を, 利用されるデバイスや Home GW の GPU 等のヘテロハードウェアにオフロードすることでの処理効率化を行っている. FPGA へは, [28] のアイデアで検討中である.

機能処理のオフロードのため, Tacit Computing は, ユーザが利用するアプリケーションのソースコードから, オフロードする領域を抽出して中間言語を出力し, 中間言語から導かれる実行ファイルを, 検証用マシンに配置実行し, オフロード効果を検証する. 検証を繰り返し, 適切なオフロード領域を定めたのち, Tacit Computing は, 実際にユーザに提供する本番環境に, 実行ファイルをデプロイし, サービスとして提供する.

図1を用いて, そのステップを説明する. 1-1 で, Tacit Computing は, ユーザに提供しているサービスの処理機能 (画像分析等) を特定する. 1-2 で, Tacit Computing は, 処理機能のソースコードを分析し, ループ文等の処理構造を把握する. 1-3 で, Tacit Computing は, ループ文等, GPU にオフロード可能な処理を特定し, オフロード処理に応じた中間言語を抽出する. 1-4 で, 中間言語ファイルを出力する. なお, 中間言語抽出は一度で終わりでなく, 適切オフロード領域探索のため, 実行試行して最適化するため反復される.

次に, Tacit Computing は, 検証用環境として, GPU を備えた検証用マシンに, 2-1 で, 中間言語から導かれる実行ファイルをデプロイする. 2-2 で, 配置したファイルを実行し, オフロードした際の性能を測定する. 詳細は, 後述するが, この性能測定結果を用いて, オフロードする領域をより適切にするため, 1-3 の中間言語抽出のステップに戻り, 別パターンの抽出を行い, 性能測定を試行する. 2-3 で, 最終的オフロード領域を指定したパターンを決定し, ユーザ向けの本番環境にデプロイされる (本番環境の選択には, [29]-[31] 等の選択技術が利用される). 2-4 で, 実行ファイル配置後, ユーザに性能を示すため, 性能試験項目をテストケース DB から抽出し, 抽出した性能試験を自動実施する (既存の, [32][33] 等を利用). 2-5 で, その試験結果を踏まえた, 性能等の情報をユーザに提示し, ユーザはサービス利用開始を判断する.

GPU 自動オフロードは, GPU に対して 1-3〜2-2 のステッ

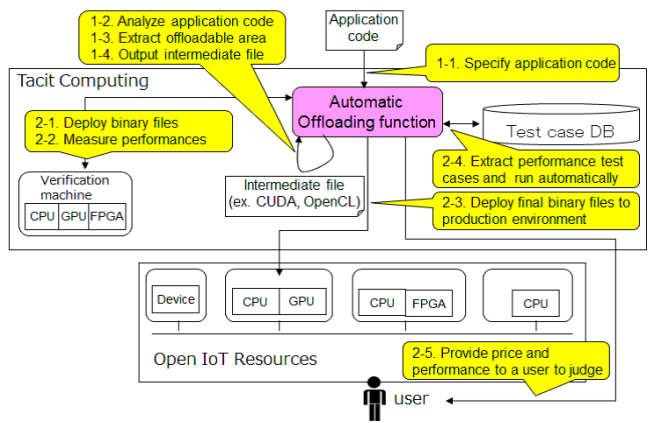

図 1 オフロード処理ステップ

プを繰り返し，最終的に 2-3 でデプロイするオフロードコードを得るための処理である．

2 節で記載の通り，高速化には適切な並列処理が必要である．特に，GPU を使う場合は，CPU と GPU 間のメモリ転送のため，データサイズやループ回数が多くないと性能が出ないことが多い．また，メモリプロセスの持ち方やメモリデータ転送のタイミング等により，並列高速化できる個々のループ文の組み合わせが，最速とならない場合等がある．

そこで, [20] では，CPU 向け汎用プログラムから，自動で適切なオフロード領域を抽出するため，並列可能ループ文群に対して遺伝的アルゴリズム（GA）[34] を用いて検証環境で性能検証試行を反復し適切な領域を探索している．

しかし, [20] では，オフロードに適切なループ文の探索はできるが，ループ毎に CPU と GPU のデータ転送が発生する場合等効率的でないことがある事や，オフロードしても高速にならない場合もある等，高速化アプリケーションが限られていた．

### 3.2 データ転送回数の低減手法

OpenACC 等の仕様では，GPU での並列処理を指定する指示行に加えて，CPU から GPU へのデータ転送やその逆を明示的に指定する指示行が定義されている．

そこで，本稿では，非効率なデータ転送を低減するため，明示的指示行を用いたデータ転送指定を，GA での並列処理の抽出と合わせて行う事を提案する．提案方式は，GA で生成された各個体について，ループ文の中で利用される変数データの参照関係を分析し，ループ毎に毎回データ転送するのでなくループ外でデータ転送して良いデータについては，ループ外でのデータ転送を明示的に指定する．

・CPU から GPU へのデータ転送．

CPU プログラム側で設定，定義した変数と GPU プログラム側で参照する変数が重なる場合は，CPU から GPU へのデータ転送が必要として，データ転送指定を行う．データ転送を指定する位置は，GPU 処理するループ文かそれより上位のループ文で，該当変数の設定，定義を含まない最上位のループとする．データ転送指示行の挿入位置は，ループの直前に行う．

・GPU から CPU へのデータ転送．

GPU プログラム側で設定した変数と CPU プログラム側で参照，設定，定義する変数が重なる場合は，GPU から CPU へのデータ転送が必要として，データ転送指定を行う．データ転送を指定する位置は，GPU 処理するループ文か，それより上位のループ文で，該当変数の参照，設定，定義を含まない最上位のループとする．データ転送指示行の挿入位置は，ループの直前に行う．ここで，設定まで含めるのは，その設定が if 文等で実行されたり，されなかったりするケースを考慮するためである．

このように，出来るだけ上位のループでデータ転送を一括して行うように，データ転送を明示的に指示することで，ループ毎に毎回データを転送する非効率な転送を避けることができる．

### 3.3 提案の GPU オフロード処理方式

これらの手法を用いても，GPU オフロードに向いていないプログラムも存在する．効果的な GPU オフロードには，オフロードする処理のループ回数が多いことが必要である．そこで，本格的なオフロード処理探索の前段階として，プロファイリングツールを用いて，ループ回数を調査することを提案する．プロファイリングツールは各行の実行回数を調査できるため，例えば，1000 万回以上のループを持つプログラムをオフロード処理探索の対象とする等，事前に振り分けることができる．

これらを備えた提案方式を図 2 を用いて説明する．

提案方式は，まず，オフロード処理部を探索するアプリケーションを分析し，for．do, while 等のループ文を把握する．次に，サンプル処理を実施し，プロファイリングツールを用いて，各ループ文のループ回数を調査し，一定の値以上のループがあるかで，オフロード処理部探索を本格的に行うかの判定を行う．

探索を本格的に行うと決まった際は，GA の処理に入る．初期化ステップでは，アプリケーションコードの全ループ文の並列可否をチェックした後，並列可能ループ文を GPU 処理する場合は 1，しない場合は 0 として遺伝子配列にマッピングする．遺伝子は指定の個体数が準備されるが，遺伝子の各値にはランダムに 1, 0 の割り当てをする．ここで，遺伝子に該当するコードでは，GPU 処理すると指定されたループ文内の変数データ参照関係から，データ転送の明示的指示を追加する．

評価ステップでは，遺伝子に該当するコードをコンパイルして検証用マシンにデプロイして実行し，ベンチマーク性能測定を行う．性能が良いパターンの遺伝子の適合度を高くする．遺伝子に該当するコードは前述のように，並列処理指示行とデータ転送指示行が挿入されている．選択ステップでは，適合度に基づいて，高適合度の遺伝子を，指定の個体数選択する．提案方式では，適合度に応じたルーレット選択及び最高適合度遺伝子のエリート選択を行う．交叉ステップでは，一定の交叉率 Pc で，選択された個体間で一部の遺伝子をある一点で交換し，子の個体を作成する．突然変異ステップでは，一定の突然変異率 Pm で，個体の遺伝子の各値を 0 から 1 または 1 から 0 に変更する．

突然変異ステップまで終わり，次の世代の遺伝子が指定個体数作成されると，初期化ステップと同様に，データ転送の明示的指示を追加し，評価，選択，交叉，突然変異ステップを繰り返す．最後に終了判定ステップでは，指定の世代数，繰り返し

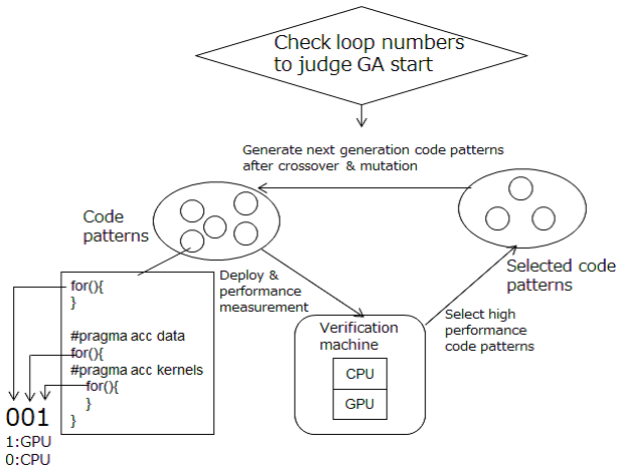

図 2　GPU オフロード探索のイメージ

を行った後に処理を終了し，最高適合度の遺伝子を解とする．

## 4. 実　　　装

3 節提案技術の有効性を確認するための実装を説明する．本稿では，GPU 自動オフロードの有効性確認が目的であるため，対象アプリケーションは C/C++ 言語のアプリケーションとし，GPU 処理自体は市中の PGI コンパイラを用いる．

GPU 処理は，PGI コンパイラにより行う．PGI コンパイラは OpenACC を解釈する C/C++/Fortran 向けコンパイラであり，for 文等の並列可能処理部を，OpenACC のディレクティブ #pragma acc kernels で指定することにより，GPU 向けバイトコードを抽出し，実行により GPU オフロードを可能としている．更に，for 文内のデータ同士に依存性があり並列処理できない処理やネストの for 文の異なる複数の階層を指定されている場合等の際に，エラーを出す．合わせて，#pragma acc data copyin/copyout/copy 等のディレクティブにより，明示的なデータ転送の指示が可能である．

実装の動作概要を説明する．実装は Perl 5 と Python 3.7 で行い，以下の処理を行う．Perl は GA の処理を，Python は構文解析を行う．

実装は，C/C++ アプリケーションの利用依頼があると，まず，C/C++ アプリケーションのコードを解析して，for 文を発見するとともに，for 文内で使われる変数データ等の，プログラム構造を把握する．構文解析には，LLVM/Clang の構文解析ライブラリ (libClang の python binding) [35] を使う．

実装は最初に，そのアプリケーションが GPU オフロード効果があるかの見込みを得るため，ベンチマークを実行し，構文解析で把握した for 文のループ回数を把握する．ループ回数把握には，GNU カバレッジの gcov [36] を用いる (gprof [37] でもよい)．テストでは，1000 万回以上のループ回数を持つアプリケーションのみ対象とするようにしたが，この値は変更可能である．

CPU 向け汎用アプリケーションは，並列化を想定して実装されているわけではない．そのため，まず，GPU 処理自体が不可な for 文は排除する必要がある．そこで，各 for 文一つずつに対して，並列処理の #pragma acc kernels ディレクティブ挿入を試行し，コンパイル時にエラーが出るかの判定を行う．コンパイルエラーは処理対象外とし，#pragma acc kernels ディレクティブは挿入しない．

ここで，並列処理してもエラーが出ないループ文の数が a の場合，a が遺伝子長となる．遺伝子の 1 は並列処理ディレクティブ有，0 は無に対応させ，長さ a の遺伝子に，アプリケーションコードをマッピングする．

次に，初期値として，指定個体数の遺伝子配列を準備する．遺伝子の各値は，0 と 1 をランダムに割当てて作成する．準備された遺伝子配列に応じて，遺伝子の値が 1 の場合は並列処理を指定するディレクティブ #pragma acc kernels を C/C++ コードに挿入する．この段階で，ある遺伝子に該当するコードの中で，GPU で処理させる部分が決まる．先ほど Clang で解析した，for 文内の変数データの参照関係を元に，3 節のルールに基づいて，CPU から GPU へのデータ転送，その逆の場合のディレクティブ指定を行う．具体的には，CPU から GPU へのデータ転送が必要な変数は，#pragma acc data copyin，GPU から CPU へのデータ転送が必要な変数は，#pragma acc data copyout で指定する．同じ変数に関して，copyin と copyout が重なる場合は，#pragma acc data copy で纏め，記述をシンプルにする．

並列処理及びデータ転送のディレクティブを挿入された C/C++ コードを，GPU を備えたマシン上の PGI コンパイラでコンパイルを行う．コンパイルした実行ファイルをデプロイし，ベンチマークツールで性能を測定する．

全個体数に対して，ベンチマーク性能測定後，ベンチマーク処理時間に応じて，各遺伝子配列の適合度を設定する．設定された適合度に応じて，残す個体の選択を行う．選択された個体に対して，交叉処理，突然変異処理，そのままコピー処理の GA 処理を行い，次世代の個体群を作成する．

次世代の個体に対して，ディレクティブ挿入，コンパイル，性能測定，適合度設定，選択，交叉，突然変異処理を行う．ここで，GA 処理の中で，以前と同じパターンの遺伝子が生じた場合は，その個体についてはコンパイル，性能測定をせず，以前と同じ測定値を用いる．指定世代数の GA 処理終了後，最高性能の遺伝子配列に該当する，ディレクティブ付き C/C++ コードを解とする．

本実装を用いて，[20] では高速化できなかった，ヤコビ法を用いてラプラス方程式を解くサンプルアプリケーション [38] の 13.5 倍の高速化を確認した．

## 5. 評　　　価

### 5.1 評価条件

#### 5.1.1 評価対象

評価対象は，IoT で多くのユーザが利用すると想定される，深層学習を用いた画像分析を行う Darknet [39] とする．Darknet は for 文が 100 以上含まれる CPU 向けの大規模なアプリケーションである．

Darknet は，C 言語で記述されたニューラルネットフレー

— 4 —

ムワークであり，画像処理以外にも，classification, detection 等様々な処理が出来るが，今回は，画像の基本処理である，オブジェクト検知や画像加工の高速化を確認する．画像処理は，様々な種類があり GPU 向けライブラリも多いが，1 つの例として C で利用可能な CPU 向けの darknet で検証する．Open IoT のサービスでカメラ映像からの自動監視や見守り等を考えた際に，画像処理は必要であり，高速化時の市場インパクトは大きい．また，Axis カメラ等，GPU を備えたカメラ等も市場に出つつある．チューニングするためのベンチマークとしては，Darknet に備え付けのサンプルアプリで detection と nightmare 処理を利用する．

### 5.1.2 評価手法

評価では，GA の各世代，各個体に対して，Darknet に対するベンチマークで性能を測定する．GA の世代変化を通じて，ベンチマーク性能が変化することを確認する．指定世代数の試行が終わった際の最高性能コードパターンが，オフロード探索の解となる．

実行する GA の，パラメータ，条件は以下で行う．

遺伝子長：並列可能ループ文数（Darknet は 75. for 文自体は，171 ある．）

個体数 M：遺伝子長以下とする（今回は 30）

世代数 T：遺伝子長以下とする（今回は 20）

適合度：$(処理時間)^{-1/2}$　処理時間が短い程，高適合度になる．また，(-1/2) 乗とすることで，処理時間が短い特定の個体の適合度が高くなり過ぎて，探索範囲が狭くなるのを防ぐ．また，性能測定が一定時間（3 分）で終わらない場合はタイムアウトさせ，処理時間 1000 秒として適合度計算する．タイムアウト時間は性能測定特性に応じて変更させる．なお，CPU のみの場合，サンプルデータを図 3 の評価環境で処理すると，Darknet は 108.28 秒である．

選択：ルーレット選択．ただし，世代での最高適合度遺伝子は交叉も突然変異もせず次世代に保存するエリート保存も合わせて行う．

交叉率 Pc：0.9

突然変異率 Pm：0.05

### 5.1.3 評価環境

利用する GPU として NVIDIA Quadro K5200 を備えた物理マシンを検証に用いる．PGI コンパイラはコミュニティ版の 17.10，CUDA Toolkit は 9.1 を用いる．評価環境とスペックを図 3 に示す．

### 5.2 性能結果

多くのユーザが IoT で使うアプリケーションとして，Darknet の高速化を確認した．GA の特性上，毎回同じ回数で収束するわけではない．そのため，それぞれ 5 回ずつ試行を行い，グラフは 5 回の中で特に収束が速くも遅くもない典型的な試行を選んでいる．

図 4 に，Darknet の，各世代個体の最高性能と GA の世代数をグラフにとる．性能は CPU のみで処理の場合との比で示している．図 4 より，20 世代の GA の中で，性能が向上しているのが分かり，全て 0（全 CPU 処理）の遺伝子では 108.28 秒

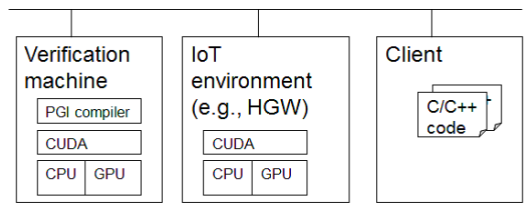

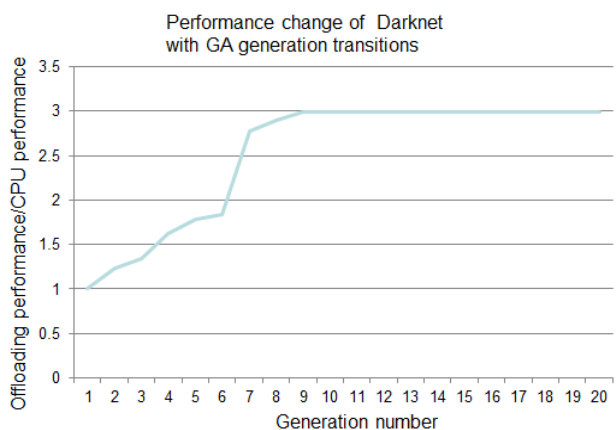

図 3　性能測定環境

図 4　GA 世代数に伴う Darknet の性能変化

だったのが，9 世代目で 36.28 秒で処理し 3 倍の性能が実現出来ていることが分かる．また，GA の各遺伝子評価の一回の試行はタイムアウト含めて平均 3 分以下であり，20 世代トータルで最大一日程度かかるはずだが，GA の中で適応度が高い同じ遺伝子パターンが生じるケースが多いこともあり，8 時間以内でオフロード抽出処理は終わった．

### 5.3 考察

以前の研究である [20] では，multiresolution analysis [40] 等に対して，GA を用いたオフロード抽出手法を用いて，数十倍の高速化を実現した．これに対して，今回，Darknet の高速化は 3 倍に留まっている．この理由としては，今回対象としているアプリケーションは，for 文が 100 以上の大規模なアプリケーションであり，全体の中で，高速化できる部分が限られていることが挙げられる．

GA のパラメータ検討やデータ転送処理パターンの拡大による，より高性能，より短時間の解発見は，今後の検討課題である．

## 6. まとめ

本稿では，Open IoT 環境で，GPU をアプリケーションで有効活用するための，Open IoT プラットフォームの要素技術として，GPU 自動オフロード技術を提案した．

提案技術は，デプロイするアプリケーションのコードから，

ループ文とループ文内のデータ参照関係を検出し，CPU-GPU間のメモリデータ転送を一括化できる部分は一括化するとともに，ループ文をオフロードするかどうかを，実際の性能結果に基づいた GA で試行探索し，適切なオフロード部を抽出する．C/C++言語のアプリケーションを対象に，PGI コンパイラを用いて提案技術を実装した．評価では，深層学習で画像分析を行う Darknet という for 文を 100 以上含む大規模のアプリケーションに対して，CPU 処理に対して 3 倍高速化できることを確認した．

文　　　献